**COMPARING CLINICAL JUDGMENT WITH *MYSURGERYRISK* ALGORITHM FOR PREOPERATIVE RISK ASSESSMENT: A PILOT STUDY**

Meghan Brennan, MD, MS[1,6,7], Sahil Puri, MS[2,7], Tezcan Ozrazgat-Baslanti, PhD[3,6], Rajendra Bhat, PhD[4,6], Zheng Feng, MS[4,6], Petar Momcilovic, PhD[5,6], Xiaolin Li, PhD[4,6], Daisy Zhe Wang, PhD[2,6], Azra Bihorac, MD, MS[3,6]

[1]Departments of Anesthesiology and [3]Medicine, University of Florida College of Medicine, Gainesville, Florida.

[2]Departments of Computer & Information Science & Engineering, [4]Electrical and Computer Engineering, [5]Industrial and Systems Engineering, University of Florida Herbert Wertheim College of Engineering, Gainesville, Florida.

[6]Precision and Intelligent Systems in Medicine (Prisma[P]), University of Florida, Gainesville, Florida.

[7]These authors contributed equally to this manuscript.

All research was conducted at the University of Florida.

Correspondence can be sent to: Azra Bihorac, MD, MS, FASN, FCCM, Department of Medicine, Division of Nephrology, Hypertension, & Renal Transplantation, 1600 SW Archer Road, PO Box 100224, Communicore Building, Room CG-98, Gainesville, FL 32610-0254. Telephone: (352) 273-9009; Fax: (352) 392-5465; E-mail: Azra.Bihorac@medicine.ufl.edu

Reprints will not be available from the authors.

**Conflicts of Interest and Source of Funding:** A.B. and T.O.B. were supported by RO1GM110240 from the National Institute of General Medical Sciences. T.O.B. has received a grant (97071) from Clinical and Translational Science Institute, University of Florida. This work was supported in part by the NIH/NCATS Clinical and Translational Sciences Award to the

**ABSTRACT**

**Background:** Major postoperative complications are associated with increased short and long-term mortality, increased healthcare cost, and adverse long-term consequences. The large amount of data contained in the electronic health record (EHR) creates barriers for physicians to recognize patients most at risk. We hypothesize, if presented in an optimal format, information from data-driven predictive risk algorithms for postoperative complications can improve physician risk assessment.

**Methods:** Prospective, non-randomized, interventional pilot study of twenty perioperative physicians at a quarterly academic medical center. Using 150 clinical cases we compared physicians' risk assessment before and after interaction with *MySurgeryRisk,* a validated machine-learning algorithm predicting preoperative risk for six major postoperative complications using EHR data.

**Results:** The area under the curve (AUC) of *MySurgeryRisk* algorithm ranged between 0.73 and 0.85 and was significantly higher than physicians' risk assessments (AUC between 0.47 and 0.69) for all postoperative complications except cardiovascular complications. The AUC for repeated physician's risk assessment improved by 2% to 5% for all complications with the exception of thirty-day mortality. Physicians' risk assessment for acute kidney injury and intensive care unit admission longer than 48 hours significantly improved after knowledge exchange, resulting in net reclassification improvement of 12.4% and 16%, respectively.

**Conclusions:** The validated *MySurgeryRisk* algorithm predicted postoperative complications with equal or higher accuracy than pilot cohort of physicians using available clinical preoperative data. The interaction with algorithm significantly improved physicians' risk assessment.

**INTRODUCTION**

Postoperative complications increase odds of 30-day mortality, lead to higher readmission rates, and greater resource utilization.[1-8] Accurate prediction of postoperative complications for individual patients is increasingly complex due to the need for rapid decision-making coupled with the constant influx of dynamic physiologic data in electronic health records (EHR).

Early warning systems such as the Modified Early Warning Score or Rothman Index, aide in rapid screening and identification of patients at risk for clinical worsening.[9, 10] Both utilize easily accessible data, are straightforward to calculate, and may be easily integrated into the EHR. These highly sensitive scores are designed to alert health care providers to all at-risk patients, but often have high false positive rates and are not specific to postoperative complications.[9, 10] Other widely validated risk scores for surgical patients such as the National Surgical Quality Improvement score and Physiological and Operative Severity Score for the enUmeration of Mortality and morbidity score provide risk stratifications for selected postoperative complications.[11, 12] The clinical use of these tools can be limited, as they require elaborate data collection and calculations.[13] Few risk scores, more specific to postoperative complications, are automatically integrated into the EHR. Interestingly studies comparing how physicians' clinical judgment compares to these risk models for predicting surgical complications are lacking.

We have developed and validated a machine learning algorithm *MySurgeryRisk* that predicts preoperative risk for eight major postoperative complications using EHR data. The algorithm is implemented in real-time in the intelligent autonomous perioperative platform developed by our group.[14-16] This platform resides in a secure environment and in real time integrates and transforms EHR data, runs predictive algorithms, produces interactive interface with outputs for physicians, inputs their feedback and prospectively collects data for the future retraining of the prediction models. The interactive interface of the platform displays risk for eight complications together with clinical variables influencing risk of that specific postoperative complication for the given clinical data.

In this prospective pilot study we compared the accuracy of risk prediction between physicians and the algorithm and tested the hypothesis that physicians will gain knowledge from interaction with the algorithm and improve accuracy of their risk assessment.

**MATERIALS AND METHODS**

*Study Design*

The University of Florida (UF) Institutional Review Board and Privacy Office approved this study (#2013-U-1338, #5-2009). This was a prospective, non-randomized, interventional pilot study of 20 intensivists with anesthesiology, medicine and surgical training who worked in surgical intensive care units at a single academic quaternary care institution. Physicians evaluated the risk of six major postoperative complications for 150 new patient cases selected from the large retrospective longitudinal cohort of adult patients, age 18 years or older, admitted for greater than 24 hours following any type of inpatient operative procedure from the years 2000 to 2010. [14, 15, 24] We selected a balanced mix of cases to assure that each complications was sufficiently represented. Physicians performed risk assessment before and after seeing the risk scores calculated for the same cases using previously validated *MySurgeryRisk* algorithm.[14-16] Each of the 150 cases was treated as an independent observation while each physicians served as their own control. Study had 80% power to detect at least 10% difference between algorithm and physician risk assessments while assuming a standard deviation of 10%.

*MySurgeryRisk Algorithm*

*MySurgeryRisk* algorithm[14-16] is validated machine-learning algorithm that predicts preoperative risk for major postoperative complications using EHR data. The algorithm is implemented in real-time in the intelligent autonomous perioperative platform developed by our group. [14-16] The platform resides on a high-performance computer maintained by UF Health information technology team in a secure environment and in real time autonomously integrates and transforms EHR data, runs predictive algorithms, produces interactive interface with outputs for physicians, inputs their feedback and prospectively collects data for the future retraining of the prediction models.[14]

For this study we have developed an interactive interface mimicking the intelligent perioperative platform that could be used with retrospective data. The interface was used to calculate and display *MySurgeryRisk* scores for six postoperative complications (30-day mortality, need for mechanical ventilation > 48 hours, cardiovascular complications, AKI, sepsis and ICU stay > 48 hours) using EHR data for 150 new patient cases. The interactive portion of the interface allowed physicians to view EHR data in a form of clinical vignette and to provide risk assessment based on their clinical judgment for each of the six complications both prior and after seeing the results of *MySurgeryRisk* algorithm.

*Physicians' Risk Assessment*

Physicians' testing took place in one-on-one sessions using personal laptops to access the interface. At the enrollment, we evaluated each physician's decision-making and numeracy with previously validated Cognitive Reflection Test (CRT) and numeracy assessment test, [21-23] respectively. The CRT consists of three questions and was developed and validated against other cognitive measures.[23] A lower score on this assessment indicates a non-reflective thinker with more impulsive decision making preference and strong reliance on intuition while higher score indicates a reflective thinker with more cautious decision making preference and less reliance on intuition.[23] During testing session, each physician evaluated 8 to 10 clinical vignettes developed from the EHR data that was used for *MySurgeryRisk* algorithm prediction. After providing their risk assessment for each case physicians were provided with the risk scores calculated by *MySurgeryRisk* algorithm. Each risk score was accompanied with the explanations of the top features that contributed the most to the calculated risk. Following this interaction physicians were asked to provide their risk assessment again. At the end of session physicians critically evaluated the interface, specifically presentation of case information, usability, and provided unstructured feedback.

*Statistical Analysis*

We calculated area under the receiver operating characteristic curve (AUC) to test performance of *MySurgeryRisk* algorithm and physicians for predicting each of the six complications separately. We assessed AUC for initial and repeated physician risk assessments after reviewing *MySurgeryRisk s*cores and compared the change using the DeLong test.[17] Net reclassification improvement (NRI) was calculated to measure the improvement in physicians' risk assessment performed after reviewing *MySurgeryRisk s*cores in comparison to the initial assessment.[18] For cases of patients who developed a complication for which risk was predicted (termed "events"), we evaluated the change in physicians' risk assessment among cases where they initially underestimated risk compared to the algorithm using the Wilcoxon signed rank sum test. We used the same methodology to examine change in physician risk assessment among cases of patients who did not develop a complication for which risk was predicted (termed "non-events") where physicians initially overestimated risk. All statistical analyses were performed using SAS software (v.9.3, Cary, N.C.) and R software (v 3.4.0, https://www.r-project.org/).

**RESULTS**

**Comparison between Physicians' Initial Risk Assessment and *MySurgeryRisk* Algorithm**

Twenty physicians provided risk assessment scores for six postoperative complications for 150 patient cases. Fifteen were attending physicians, with an average of thirteen years of experience. Ninety percent obtained a high score on the numeracy assessment. The majority, 70%, scored in the intermediate range for the decision-making style, with only 15% of physicians scoring in impulsive and other 15% in reflective decision makers range (Table 1).

Among 150 patient cases, prevalence of postoperative complications ranged between 16% for 30-day mortality and 49%, for ICU admission > 48 hours (Table 2). *MySurgeryRisk* algorithm was more accurate in predicting risk for complications compared to physicians (Table 2). Physicians demonstrated an AUC between 0.47 and 0.69 in predicting risk for postoperative complications, while *MySurgeryRisk* algorithm's AUC ranged from 0.64 to 0.85. The differences in AUCs between the algorithm and initial physicians' risk assessment were statistically significant for all complications except cardiovascular (Table 2). Physicians were most likely to underestimate risk of ICU stay and AKI and most likely to overestimate risk of mortality, CV complications, and severe sepsis.

**Change in Physicians' Risk Assessment after interaction with *MySurgeryRisk* Algorithm**

To assess whether physicians changed their risk assessment after reviewing *MySurgeryRisk* algorithm results we compared their repeated and initial risk assessments. For each complications separately we stratified cases as events (cases for which postoperative complication occurred) and non-events (cases for which postoperative complication did not occur). For all postoperative complication events and non-events, the majority of physicians responded to interaction with the algorithm by appropriately increasing or decreasing score, respectively (Table 3). At initial risk assessment, physicians tended to overestimate risk of non-events for postoperative complications and underestimate risks of events. Their estimates improved after interaction with the algorithm (Table 3). The calculated net reclassification improvement (net percentages of correctly reclassified cases of events or non-events after interaction with the algorithm) showed statistically significant improvement for AKI and ICU admission > 48 hours with values of 12.4% and 16%, respectively (Table 4). The AUC for repeated physician's risk assessment improved by 2% to 5% for all complications with the exception of thirty-day mortality. The improvement in AUC for predicting CV complications before and after their interaction was only one that was statistically significant, increasing by 5% (Table 4).

Although study size was too small for formal comparison, decision-making attitudes as classified by the CRT appear to play a role in physician interaction with the algorithm. Physicians scoring as intuitive decision makers on the CRT did not change their risk assessments as much as physicians scoring as reflective decision makers did. This was most noticeable in cases in which they overestimated risks of non-events.

**Cases with Extreme Disagreement**

We considered cases to have extreme disagreement if the algorithm and physician's risk scores differed by at least 40 points (on the scale 0 to 100) resulting in a change from low to high risk category or opposite. We considered any patient to be in the high risk category if the risk score was above previously reported prevalence threshold for each postoperative complication.[14-16] Extreme disagreement occurred among 8% (75/900) of all case comparisons across six complications and was more common in cases of non-events (47/292, 16%) compared to event cases (28/608, 5%). For the majority of disagreement cases patients who did not develop complication were assigned my physician to high risk category while *MySurgeryRisk* algorithm correctly assigned them to low risk. For patients who developed complication but physician assigned them to low risk category, the algorithm correctly assigned to high risk group when assessing risk for AKI, ICU admission and MV.

**Physician Evaluation of the Software**

A post-test survey, with five Likert scale and nine free response questions, was given to all participants, and half completed the evaluation. The majority answered that the software was easy to use; one respondent reported difficulty with the sliding bar. Half answered that the software helped with decision-making processes; three responded it did not help with their decision-making process and they did not change risk assessment scores. The majority listed tablet and website-based applications during clinics and ICU rounds as the best place to access the software. Two reported they would use it for counseling patients preoperatively.

**DISCUSSION**

In this pilot study we have demonstrated that validated machine-learning *MySurgeryRisk* algorithm predicted postoperative complications with equal or higher accuracy than our sample of physicians using only available clinical preoperative data. *MySurgeryRisk* is validated machine-learning algorithm that uses existing clinical data in electronic health records to predict the risk for major complications and death after surgery with high sensitivity and high specificity. Interestingly physicians were more likely than the algorithm to both underestimate risk of postoperative complication for cases where complication actually occurred and overestimate risk for cases when complication did not occur. The interaction with *MySurgeryRisk* algorithm resulted in significant improvement in physicians' risk assessment, as evidenced by improvements of both AUC and net reclassification improvement scores for the tested postoperative complications. The computational algorithm as a higher-capacity and lower-cost information processing service is a logical next step to support physician's decision making for rapid identification of patients at risk in perioperative period. *MySurgeryRisk* informs physician about important clinical variables contributing to the risk to clarify the basis for algorithm risk score and to facilitate users' confidence in using the algorithm. The algorithm is currently deployed in a clinical workflow for real-time autonomous surgical risk prediction as a part of a single-center prospective clinical trial.[16]

Physician's abilities to predict postoperative outcomes and comparison of physician to automated predictive risk scores and systems have not been studied extensively and studies have produced mixed findings.[15, 16, 21, 22, 25, 26] Among few studies comparing differential diagnosis generators, symptom checkers, and automated electrocardiogram with physicians, although algorithms improved accuracy in less acute or more "common" scenarios, in general physicians had better diagnostic accuracy.[25, 26] Studies specific to colorectal and hepatobiliary surgery, showed the surgeon's gut feeling was more accurate than POSSUM score to predict postoperative mortality.[27, 28] Ivanov et al showed that physicians tended to overestimate risk of postoperative mortality and prolonged ICU stay, in patients undergoing coronary artery bypass surgery when compared to a statistical model.[29] Detsky et al. showed that accuracy for ICU physicians' prediction of in-hospital mortality, return to home at 6 months, and 6-month cognitive function varied considerably and was only slightly better than random.[30] With improvement of computing capabilities, need to process the constantly changing large volume data contained in the EHR, and need to provide patients and families with relatively specific estimates of outcomes, predictive models and decision-making aids are going to be increasingly necessary in healthcare.[31] Our algorithm outperforms physician prediction postoperative complications in the majority of cases. Overall, physicians tended to overestimate risks of sentinel events such as thirty-day mortality and cardiovascular events and are less likely to significantly reduce risk assessment after interaction with the algorithm. It is likely because these complications are sentinel events and are associated with significant emotional, personal, and professional consequences of unrecognized risk. On the other hand, physicians were less and more perceptive for interaction with algorithm when dealing with complications that they have less knowledge about or that they may perceive as less acute, such as AKI or ICU admission.

Physicians decision-making style can influence how they perceive risk and use information from decision tools such as algorithms. In spite of small sample size we observed that physicians who scored at the extreme of cognitive reflection test reacted differently in response to interaction with the algorithm. Reasons for this are not known but studies of decision making preferences by Frederick et al suggest that this could be due to time spent on risk assessment and willingness to reassess decision making process as assessed by cognitive reflection test that measures people's ability to resist their first instinct.[23] A high score indicates a reflective thinker whose initial intuition is tempered by analysis and who takes more time to reflect on risk probabilities and information provided by the algorithm. Although our participants had high numeracy scores, it has been widely demonstrated that even those individuals are likely to make numerical mistakes on relatively simple numeric questions.[21, 22] Additionally, there may be a more optimal way to present numeric clinical data that encourages more accurate risk assessment.

Our study has several limitations. Data used for *MySurgeryRisk* algorithm, although more than sufficient in size to have well-fitting and precise models, was collected from a single center. Results may not be generalizable where patient characteristics differ dramatically. Second, our number of physician participants was small and homogeneous, making it difficult to provide significant information about physician decision-making preferences based on cognitive reflection test and numeric assessment. Third, physicians may have estimated risks more accurately if they had exposure to an increased amount of patient data.

A majority of respondents to the post-test survey found the system easy to use, helpful for decision making, and appropriate for the clinical environment. We are further refining the algorithm, allowing participants to input their own assessments into the computational algorithm to facilitate two-way knowledge transfer and allow models to "learn" from participants. We anticipate expanding our range of complications to allow for greater personalization specific to individual patients and to include the algorithm risk assessment scores into the EHR. The large prospective clinical evaluation of the algorithm in multiple real-time environments to assess algorithm and participant performance, ease of use in clinical decision-making, and potential for further reduction of postoperative complications is ongoing.[16]

Prediction of major postoperative complications is complex and multifactorial; in our study, we were able to demonstrate that novel machine-learning *MySurgeryRisk* algorithm implemented in real-time autonomous intelligent platform improved physician accuracy in predicting postoperative complications. Further studies are required with a larger physician sample and an increased number of case scenarios to verify our results. The implementation of autonomous platform with capacity for real-time analytics and communication with physicians in perioperative clinical workflow would greatly simplify and augment perioperative risk assessment and stratification of patients.

**ACKNOWLEDGMENTS**

Meghan Brennan and Sahil Puri contributed equally to the manuscript.

**Table 1.** Physician Characteristics.

| Physician characteristics | N=20 |
|---|---|
| Female gender, n (%) | 5 (25) |
| Age (years), n (%) | |
|    <=30 | 2 (10) |
|    31-40 | 10 (50) |
|    41-50 | 4 (20) |
|    >50 | 4 (20) |
| Attending doctor, n (%) | 15 (75) |
| Specialty, n (%) | |
|    Anesthesiology | 13 (65) |
|    Surgery | 4 (20) |
|    Emergency room | 2 (10) |
|    Medicine | 1 (5) |
| Years since graduation, mean (SD) | 13 (10) |
| High numeracy score (>=9), n (%) | 18 (90) |
| Cognitive Reflection Test score, n (%) | |
| *Measures people's ability to resist their first instinct* | |
|   Low score (0): Non-reflective thinker with unquestioning reliance on intuition | 3 (15) |
|   Intermediate score (1-2) | 14 (70) |
|   High score (3): Reflective thinker whose initial intuition is tempered by analysis | 3 (15) |

**Table 2.** Comparison between Physicians' Initial Risk Assessment and *MySurgeryRisk* Algorithm Prediction.

| Postoperative complications | Prevalence of complications among cases, n (%) | Physicians' first risk assessment AUC (95% CI) | *MySurgeryRisk* Algorithm AUC (95% CI) | p-value for difference in AUC |
|---|---|---|---|---|
| Intensive care unit admission longer than 48 hours | 74 (49) | 0.69 (0.61, 0.77) | 0.84 (0.78, 0.90) | 0.0006 |
| Acute kidney injury | 57 (38) | 0.65 (0.56, 0.74) | 0.79 (0.72, 0.87) | 0.0017 |
| Mechanical ventilation longer than 48 hours | 55 (37) | 0.66 (0.57, 0.75) | 0.85 (0.79, 0.91) | <0.0001 |
| Cardiovascular complications | 43 (29) | 0.54 (0.44, 0.65) | 0.64 (0.55, 0.73) | 0.09 |
| Severe sepsis | 39 (26) | 0.54 (0.44, 0.64) | 0.78 (0.69, 0.87) | <0.0001 |
| Thirty-day mortality | 24 (16) | 0.47 (0.36, 0.57) | 0.73 (0.64, 0.83) | <0.0001 |

Abbreviations: AUC, area under the receiver operator characteristic curve; CI, confidence interval.

**Table 3.** Change in Physicians' Risk Assessment after Interaction with *MySurgeryRisk* Algorithm.

| | Events[a] | | | Non-events[a] | | |
|---|---|---|---|---|---|---|
| Postoperative complications | Physician under-estimated risk initially, n (%) | Physician increased score, n (%) | Change in repeated risk score, mean (SD)[b] | Physician over-estimated risk initially, n (%) | Physician decreased score, n (%) | Change in repeated risk score, mean (SD) |
| Intensive care unit admission longer than 48 hours | 51/74 (69) | 24/51 (47) | 4 (12)[c] | 37/76 (49) | 31/37 (84) | −13 (13)[c] |
| Acute kidney injury | 37/57 (65) | 26/37 (70) | 6 (13)[c] | 44/93 (47) | 28/44 (64) | −8 (12)[c] |
| Mechanical ventilation longer than 48 hours | 31/55 (56) | 16/31 (52) | 7 (15)[c] | 63/95 (66) | 43/63 (68) | -10 (17)[c] |
| Cardiovascular complications | 17/43 (39) | 12/17 (71) | 3 (4)[c] | 78/107 (73) | 56/78 (72) | −8 (12)[c] |
| Severe sepsis | 21/39 (54) | 16/21 (76) | 6 (8)[c] | 76/111 (68) | 52/76 (68) | −7 (12)[c] |
| 30-day mortality | 13/24 (54) | 5/13 (38) | -1 (4) | 97/126 (77) | 65/97 (67) | −5 (13)[c] |

[a]Events are clinical cases for which predicted complication has occurred and non-events are clinical cases for which predicted complication has not occurred.

[b]All physicians risk assessment scores were on the scale from 0 (no risk) to 100 (complete certainty of risk).

[c]p-value < 0.05. The change in repeated physicians' risk assessment scores after interaction with *MySurgeryRisk* Algorithm was tested using Wilcoxon signed rank sum test.

**Table 4.** Comparison between Initial and Repeated Physicians' Risk Assessment after interaction with *MySurgeryRisk* Algorithm.

| Postoperative complications | Physicians' initial risk assessment AUC (95% CI) | Physicians' risk re-assessment AUC (95% CI) | p-value for difference in AUC | Net reclassification improvement, % (95% CI) |
|---|---|---|---|---|
| Intensive care unit admission longer than 48 hours | 0.69 (0.61, 0.77) | 0.71 (0.62, 0.79) | 0.452 | 16.0 (3.0, 29.6) [b] |
| Acute kidney injury | 0.65 (0.56, 0.74) | 0.69 (0.60, 0.77) | 0.064 | 12.4 (1.0, 23.8) [b] |
| Mechanical ventilation longer than 48 hours | 0.66 (0.57, 0.75) | 0.70 (0.61, 0.80) | 0.074 | 0.8 (-10.9, 9.3) |
| Cardiovascular complications | 0.54 (0.44, 0.65) | 0.59 (0.49, 0.69) | 0.039[a] | 5.1 (-2.9, 13.1) |
| Severe sepsis | 0.54 (0.44, 0.64) | 0.59 (0.50, 0.69) | 0.063 | 7.8 (-5.9, 21.6) |
| Thirty-day mortality | 0.47 (0.36, 0.57) | 0.49 (0.39, 0.60) | 0.276 | -1.0 (-11.6, 9.6) |

Abbreviations: AUC, area under the receiver operator characteristic curve; CI, confidence interval.

[a] p-value < 0.05. The change in AUC for repeated physicians' risk assessment after interaction with *MySurgeryRisk* Algorithm was tested using the DeLong test.

[b] p-value < 0.05.

**REFERENCES**

1. Khuri, S.F., et al., *Determinants of long-term survival after major surgery and the adverse effect of postoperative complications.* Ann Surg, 2005. **242**(3): p. 326-41; discussion 341-3.

2. Silber, J.H., et al., *Changes in prognosis after the first postoperative complication.* Med Care, 2005. **43**(2): p. 122-31.

3. Hobson, C., et al., *Cost and Mortality Associated With Postoperative Acute Kidney Injury.* Ann Surg, 2015. **261**(6): p. 1207-14.

4. Hobson, C., G. Singhania, and A. Bihorac, *Acute Kidney Injury in the Surgical Patient.* Crit Care Clin, 2015. **31**(4): p. 705-23.

5. Vogel, T.R., et al., *Postoperative sepsis in the United States.* Ann Surg, 2010. **252**(6): p. 1065-71.

6. Lagu, T., et al., *Hospitalizations, costs, and outcomes of severe sepsis in the United States 2003 to 2007.* Crit Care Med, 2012. **40**(3): p. 754-61.

7. Bihorac, A., *Acute Kidney Injury in the Surgical Patient: Recognition and Attribution.* Nephron, 2015. **131**(2): p. 118-22.

8. Bihorac, A., et al., *National surgical quality improvement program underestimates the risk associated with mild and moderate postoperative acute kidney injury.* Crit Care Med, 2013. **41**(11): p. 2570-83.

9. Tepas, J.J., et al., *Automated analysis of electronic medical record data reflects the pathophysiology of operative complications.* Surgery, 2013. **154**(4): p. 918-924.

10. Hollis, R.H., et al., *A Role for the Early Warning Score in Early Identification of Critical Postoperative Complications.* Ann Surg, 2016. **263**(5): p. 918-23.

11. Copeland, G.P., D. Jones, and M. Walters, *POSSUM: a scoring system for surgical audit.* Br J Surg, 1991. **78**(3): p. 355-60.

12. Gawande, A.A., et al., *An Apgar score for surgery.* J Am Coll Surg, 2007. **204**(2): p. 201-8.

13. *ACS NSQIP Surgical Risk Calculator*. 2007-2018; Available from: https://riskcalculator.facs.org/RiskCalculator/.

14. Bihorac, A., et al., *MySurgeryRisk: Development and Validation of a Machine-learning Risk Algorithm for Major Complications and Death After Surgery.* Ann Surg, 2018.

15. Thottakkara, P., et al., *Application of Machine Learning Techniques to High-Dimensional Clinical Data to Forecast Postoperative Complications.* PLoS One, 2016. **11**(5): p. e0155705.

16. Feng Z, R.B.R., X Yuan, D Freeman, T Baslanti, A Bihorac, X Li *Intelligent Perioperative System: Towards Real-time Big Data Analytics in Surgery Risk Assessment.* arXiv, 2017.

17. Delong, E.R., D.M. Delong, and D.I. Clarkepearson, *Comparing the Areas under 2 or More Correlated Receiver Operating Characteristic Curves - a Nonparametric Approach.* Biometrics, 1988. **44**(3): p. 837-845.

18. Pencina, M.J., R.B. D'Agostino, and E.W. Steyerberg, *Extensions of net reclassification improvement calculations to measure usefulness of new biomarkers.* Statistics in Medicine, 2011. **30**(1): p. 11-21.

19. *Agency for Healthcare Research and Quality: Patient Safety Indicators: Technical Specifications.* 2010 [cited 2017 January 25]; Available from: http://www.qualityindicators.ahrq.gov/modules/psi_overview.aspx.

20. Elixhauser, A., et al., *Comorbidity measures for use with administrative data.* Medical Care, 1998. **36**(1): p. 8-27.

21. Lipkus, I.M., G. Samsa, and B.K. Rimer, *General performance on a numeracy scale among highly educated samples.* Medical Decision Making, 2001. **21**(1): p. 37-44.

22. Schwartz, L.M., et al., *The role of numeracy in understanding the benefit of screening mammography.* Annals of Internal Medicine, 1997. **127**(11): p. 966-972.

23. Frederick, S., *Cognitive reflection and decision making.* Journal of Economic Perspectives, 2005. **19**(4): p. 25-42.

24. *United States Census Bureau: American FactFinder* 2010 [cited 2017 January 25]; Available from: https://factfinder.census.gov/faces/nav/jsf/pages/index.xhtml.

25. Poon, K., P.M. Okin, and P. Kligfield, *Diagnostic performance of a computer-based ECG rhythm algorithm.* Journal of Electrocardiology, 2005. **38**(3): p. 235-238.

26. Semigran, H.L., et al., *Comparison of Physician and Computer Diagnostic Accuracy.* Jama Internal Medicine, 2016. **176**(12): p. 1860-1861.

27. Markus, P.M., et al., *Predicting postoperative morbidity by clinical assessment.* British Journal of Surgery, 2005. **92**(1): p. 101-106.

28. Hartley, M.N. and P.M. Sagar, *The surgeon's 'gut feeling' as a predictor of post-operative outcome.* Ann R Coll Surg Engl, 1994. **76**(6 Suppl): p. 277-8.

29. Ivanov, J., et al., *Predictive accuracy study: comparing a statistical model to clinicians' estimates of outcomes after coronary bypass surgery.* Ann Thorac Surg, 2000. **70**(1): p. 162-8.

30. Detsky, M.E., et al., *Discriminative Accuracy of Physician and Nurse Predictions for Survival and Functional Outcomes 6 Months After an ICU Admission.* JAMA, 2017. **317**(21): p. 2187-2195.

31. Liao, L. and D.B. Mark, *Clinical prediction models: Are we building better mousetraps?* Journal of the American College of Cardiology, 2003. **42**(5): p. 851-853.